\documentstyle[11pt,newpasp,twoside,epsf]{article}
\def\edcomment#1{\iffalse\marginpar{\raggedright\sl#1\/}\else\relax\fi}
\reversemarginpar

\begin{document}
\title{On the nature of the Extremely Red Galaxies}
\author{Filippo Mannucci}
\affil{CAISMI - CNR, Largo E. fermi 5, 50125 Firenze, Italia}

\begin{abstract}
We present a method based on two broad-band colors to investigate
the nature of the Extremely Red Objects (EROs), i.e., the 
galaxies selected to have very red optical-to-infrared colors.
Dusty starburst and old ellipticals at redshifts between 1 and 2 appear to 
occupy two different regions of the J-K vs. I-K color diagram, allowing for an
easy classification. This diagnostic was applied to a complete sample of 57
EROs: the two populations are found to be present in the sample in similar
abundances. The cosmic star formation density in the dusty starbursts 
is found to be of the order of that in the Lyman-Break Galaxies (LBG).
\end{abstract}

\section{Introduction}

EROs proved to be a very difficult population to study. They are too
faint both in the optical and in the near-IR to obtain large number of spectra
with enough signal-to-noise ratio to allow for a secure classification.
Even the most recent deep spectroscopic surveys (e.g., Cimatti et al, 2002)
can only reach K$\sim$19.2 and even in this case the number of unidentified
objects is similar to the number of objects of each class.
We have therefore searched for a diagnostics that could be
applied to large sample of EROs without large amounts of telescope time.
We noticed (Pozzetti \& Mannucci, 2000) that the optical and near-IR 
Spectral Energy Distribution (SED) of a dusty object is expected to be
swallower than that of an old elliptical because in the
latter case the red color is given by the sharp 4000\AA\ break (see fig. 1).
As a consequence, dusty starburst with z$<$2 are expected to have
redder J-K colors than old ellipticals. The positions of the two populations in
the J-K vs. I-K (or R-K) color diagram were estimated by a large number of
synthetic galaxy models.
The expected difference in the J-K color is about 0.3 mag, even if
for some models the difference reduces to zero. The derived diagnostic
correctly reproduces all the objects with known nature (see fig. 2).
A good photometry is needed to obtain reliable results as
no empty region is present between the two populations and the
relative distance is not large. 

\begin{figure}
\plotfiddle{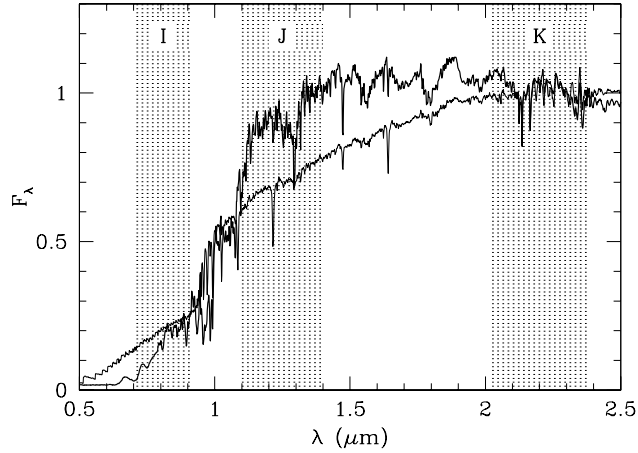}{5.5cm}{0}{45}{45}{-140}{-100}
\caption{
Representative spectra from an old elliptical (thick line) and a dusty
starburst (thin line) at z=1.5. The position of the I, J and K filters
are shown. 
It is apparent as, while the I-K colour selects both class of objects,
the J-K colour can distinguish between them.
}
\end{figure}

\begin{figure}
\plotfiddle{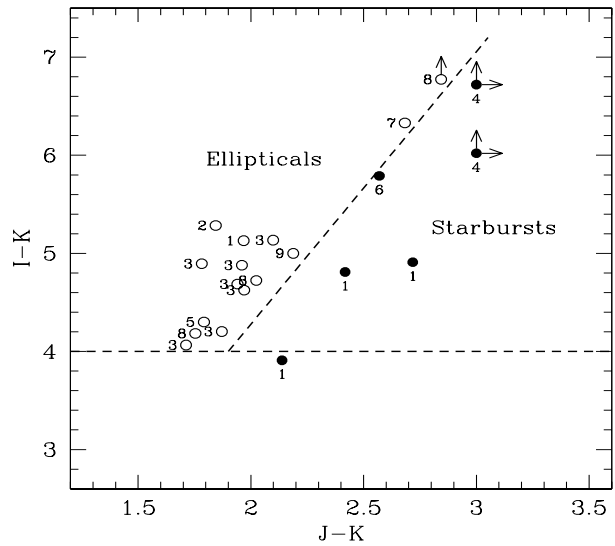}{6.5cm}{0}{45}{40}{-140}{-60}
\caption{
Comparison of the selection criterion derived from the models with 
observations.
Empty dots are elliptical galaxies, solid dots are
dusty starbursts.  
Data are labeled according to the data source (See Pozzetti \& Mannucci 2000
for details)
}
\end{figure}

We have applied this diagnostic to a complete sample of 57 EROs selected to
have K$<$20.3 and R-K$>$5.3 by Thompson et al (2000) in the field of 
a radio galaxy (see Mannucci et al., 2001, for details). 
The color-color diagram is
shown in the left panel of fig. 3, together with the separation line.
According to the models and the observations of cold dwarfs,
the objects with J-K$<$1.4 were classified as stars. Among the other objects,
21 (37\% of the sample) can be classified as starburst, the same number as
ellipticals, and 10 objects (17\% of the sample) remain unclassified.
Several objects fall near the separation line so that their
classification can be affected by the uncertainties on the colours.
The significance of the classification was estimated by the ratio of the
distance of each point from the classification line divided by the
radius of the error ellipse in the direction perpendicular to the line.
These results are shown in the right panel of fig. 3.
Eleven ellipticals and 2 starbursts have a distance from the line
below 1 sigma, therefore their classification should be considered
weak.

We have checked the effect of the uncertainties 
on the relative number of objects of the two populations.
While the statistical errors are unlikely to affect the result very much, the
systematic effect on the colors and on the position of the line can effect the
ratio between ellipticals and dusty galaxies of about 35\%.

\begin{figure}
\plotfiddle{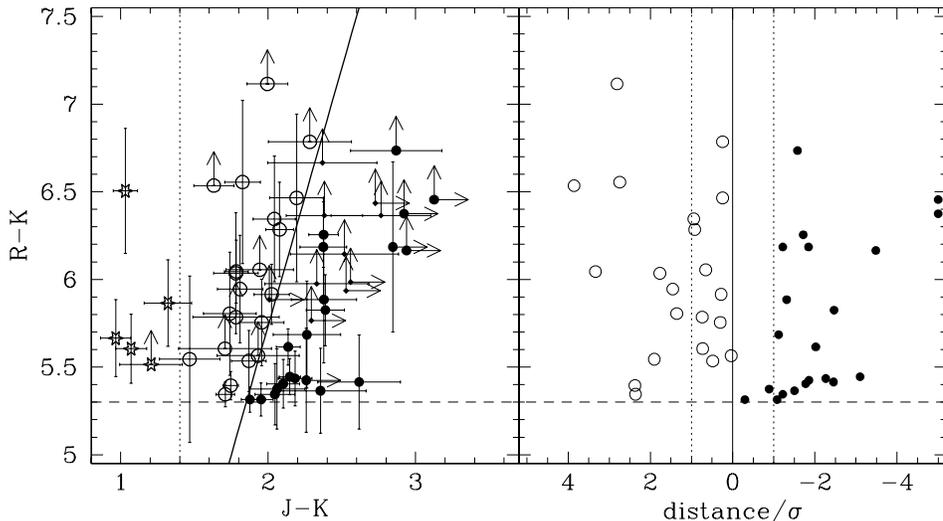}{6cm}{-90}{50}{50}{-200}{240}
\caption{
Left panel: R$-$K vs. J$-$K colour diagram of the 
selected EROs.
The thick line is the separation between ellipticals and starbursts derived
following Pozzetti \& Mannucci (2000). 
The horizontal dashed line is the colour threshold for
selection, the dotted vertical line the J$-$K colour of the separation
between  stars and galaxies.  Open circles represent objects classified 
as ellipticals, while filled circles represent the starbursts.  Stars 
are used for the stars. 
Objects with no classification are shown as small diamonds.\\
Right panel: significance of the classification as elliptical or
starburst. On the x axis is plotted the
ratio between the distance of each point from the separation line and
the radius of the error ellipse in the direction perpendicular to
the line. 
The dotted lines show the 1$\sigma$ limit of the classification.
}
\end{figure}

We have studied the morphology of the objects to check the
consistency of the classification. A correlation was found between the K-band
morphology and the classification: as expected, the objects identified as
ellipticals have regular morphology and small angular dimensions, while the
starburst objects appear to be more extended and often elongated.\\

We have derived a crude estimate of the contribution of the detected 
starburst galaxies to the cosmic star formation density. The cosmic variance
and uncertainties on the Star Formation Rate (SFR) of each galaxies 
and on the cosmic volume actually sampled make the resulting value vary
uncertain. Nevertheless a value of 0.03  M$_{\odot}$/yr was derived
for a $\Lambda$ cosmology with H$_0$=70, $\Omega_m=0.3$ and 
$\Omega_{\Lambda}=0.7$. 
This is of the same order of the value derived from the 
LBG at higher redshift. This is an indication
that the ERO technique is not only a good method to detect
high-redshift ellipticals, but also a promising way to select a
population of starburst galaxies containing a significant fraction
of star formation and having intermediate properties between
the LBG and the SCUBA objects (see fig. 4).
It is also interesting to note that the results 
on the relative number of objects and cosmic star formation history
of this work 
are in a very good agreement with those of the following 
K20 spectroscopic survey (Cimatti et al., 2002) even if the 
techniques are radically different.

\begin{figure}
\plotfiddle{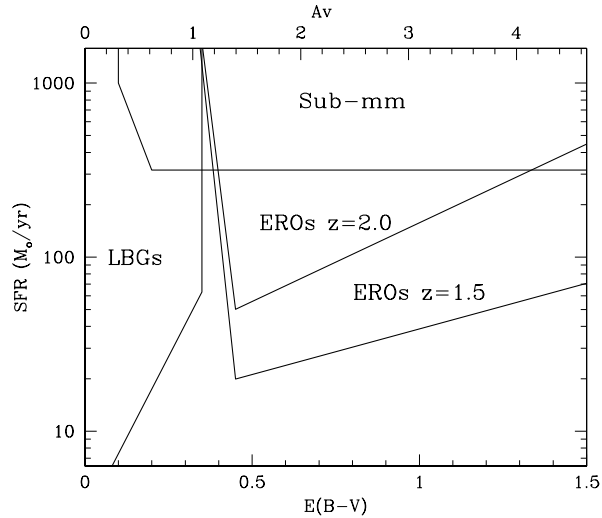}{5.6cm}{0}{40}{40}{-130}{-70}
\caption{
Regions of the SFR - extinction plane that can be selected by
the various search techniques. Sub-mm observations can currently detect
galaxies with any extinction but with very high SFRs.
LBG must have E(B$-$V) below about 0.3, but can have low SFRs. 
The ERO
technique covers a region of this plane with E(B$-$V) between 0.5 and 1 and
a star formation rate down to about 20 M$_{\odot}$/yr.  The exact 
values of the edges depend on many assumptions, such as the choice of 
telescope and instrument, the integration time, dust temperature, and 
redshift.  Therefore the limits should be regarded only as
indicative. Only methods aimed to detect the continuum are shown; 
techniques targeting emission lines are not included in the plot.
}
\end{figure}


\begin{references}
\reference Cimatti, A., et al., 2002, \aap, in press (astro-ph/0111527).
\reference Mannucci, F., et al., 2001, \mnras, in press (astro-ph/0110648)
\reference Pozzetti, L., \& Mannucci, F., 2000, \mnras, 317, L17
\end{references}
\end{document}